\documentclass[aip,rsi,amsmath,amssymb,reprint]{revtex4-1}
\usepackage{amsmath}
\usepackage{amsbsy}
\usepackage{bm}
\usepackage{graphicx}
\usepackage{mathtools}
\usepackage[colorinlistoftodos]{todonotes}
\usepackage[colorlinks=true, allcolors=blue]{hyperref}
\DeclareMathAlphabet{\mathpzc}{OT1}{pzc}{m}{it}

\begin{document}

\title{Operators in Machine Learning: Response Properties in Chemical Space}

\author{Anders S. Christensen}
\affiliation{Department of Chemistry, University of Basel, Basel, Switzerland}

\author{Felix A. Faber}%
\affiliation{Department of Chemistry, University of Basel, Basel, Switzerland}

\author{O. Anatole von Lilienfeld}
\email{anatole.vonlilienfeld@unibas.ch}
\affiliation{Department of Chemistry, University of Basel, Basel, Switzerland}

\begin{abstract}
The role of response operators is well established in quantum mechanics. 
We investigate their use for universal quantum machine learning models of response properties in molecules. 
After introducing a theoretical basis, we present and discuss numerical evidence 
based on measuring the potential energy's response with respect to atomic displacement and to electric fields. 
Prediction errors for corresponding properties, atomic forces and dipole moments, 
improve in a systematic fashion with training set size and reach high accuracy for small training sets. 
Prediction of normal modes and IR-spectra of some small molecules demonstrates the usefulness of this approach for chemistry.
\end{abstract}

\maketitle

\section{Introduction}
Time-independent electronic ground-state quantum properties can be expressed as expectation values of the electronic wave function and an operator, typically defined via the correspondence principle. 
The performance of supervised machine learning models of these quantum properties, a.k.a.~quantum machine learning (QML),~\cite{RuppPRL2012, Montavon2013,ML4Polymers_Rampi2013,Elpasolite_2016} can be conveniently assessed using learning curves which monitor the decay of the out-of-sample prediction error (deviation of predicted properties from reference for query compounds not included in training) as a function of compound training set size $N$. 
Due to the leading prediction error decaying as $a/N^b$, log-log plots have become the recommended practice in the field with $\log(a)$ and $b$ denoting the off-set and learning rate (or efficiency), respectively~\cite{vapnik1994learningcurves,StatError_Muller1996,QMLessayAnatole}. 
While in principle, supervised ML models can be generated for any cause and effect relationship, it is the very philosophy of QML that representation (and kernel function when using kernel ridge regression) be property independent~\cite{SingleKernel2015, RaghusReview2016} in the same way in which the electronic wave function and its Hamiltonian are property independent.
However, there is a select and highly relevant set of quantum properties which can 
be understood as response properties, obtained through the use of response operators and perturbation theory. 
Common examples include derivatives of the energy with respect to e.g.~the nuclear displacement or charge, an external electric field, an external magnetic field, or nuclear magnetic moments, and can efficiently be accounted for within density functional theory~\cite{gonze1995perturbationtheory,apdsmp}.
We note in passing that energy response properties also form the basis for conceptual density functional theory~\cite{parryang,Geerlings_DFTConcepts}, 
as well as computational alchemy~\cite{anatole-prl2005,anatole-jcp2009-2,CatalystSheppard2010,anatole-ijqc2013,Samuel-JCP2016,AlchemyAlisa_2016,StijnPNAS2017,Samuel2018bandgaps}. 
It has previously been observed that prediction errors of many conventional quantum machine learning models of response properties can converge relatively slowly, even for machine models that are able to achieve remarkably high accuracy for energies~\cite{Montavon2013,SingleKernel2015,googlePaper2017,FCHL,pronobis2018many}.
In this paper we investigate if the use of response operators is beneficial for deriving improved QML models which afford learning curves with lower off-sets and better learning rates.   

Maybe the most relevant quantum response property is the force exerted on each atom in the system, the first order energy derivative with respect to nuclear displacement~\cite{HF}. 
Quite recently, tremendous efforts have been made to predict atomic forcces accurately within QML models for the purpose of running {\em ab initio} quality molecular dynamics simulations at low computational cost.\cite{SumpterNoidNeuralNetworks1992,Neuralnetworks_Scheffler2004,NN_Tucker2006,Neuralnetworks_BehlerParrinello2007,bpkc2010,BartokGabor_Descriptors2013,ManzhosCarrington_NNreview2015,RampiMLQMMM,MLatoms_2015,chmiela2017machine,ANI_IsayevRoitberg2017,schutt2018schnet,grisafi2018symmetry}.
Treating the force as the first derivative of the energy is tantamount to using the gradient operator, as commonly implemented in quantum chemistry packages.
Doing so leads directly to energy conservation, a crucial property for most statistical mechanics applications, which has already also been obtained by others~\cite{CovariantKernelsSandro2016,chmiela2017machine}.
The use of response operators, however, has not yet been applied generally to generate QML models for other response properties.

Here, we extend the principle of using response operators to investigate the potential total energy and its response to a change in (i) atomic coordinates, and (ii) an external electric field, i.e.~the dipole moments. Other QML models, capable of predicting dipole moments have already been published.\cite{Montavon2013, SingleKernel2015, bing2016,SchuettTkatchenkoMueller2017,Sifain2018,nebgen2018transferable,gastegger2017machine,schutt2018quantum}
The work by Sch\"utt \textit{et al.} presents a neural network that is able to predict the dipole moment of the QM9 dataset\cite{GDB17,DataPaper2014} with very high accuracy\cite{SchuettTkatchenkoMueller2017}, by simply training on the observable, the dipole moment vector itself.
Other approaches rely on a charge model predicted from a neural network to estimate intensities in an infrared spectrum where the frequencies are obtained from a molecular dynamics simulation.\cite{gastegger2017machine,Sifain2018} 
Similarly to Sch\"utt \textit{et al.}, we propose to learn the dipole moment by training on the quantum mechanical observable directly, but in contrast 
we train a model to describe the energy for which the dipole moment can be calculated as a response property simply by taking the derivative.  
The modeling of highly accurate molecular potential energy surfaces has also been thoroughly investigated with several ML techniques, 
due to their important connection to infrared (IR) spectroscopy.\cite{Carrington2006,Carrington2006_2,Carrington2008,Krems2016}
We show how our operator formalism can lead to ML potential energy surfaces that reproduce the 
vibrational normal modes of molecules across chemical space, and even reproduces the IR spectrum of a molecule simply by using the relevant response operators with a suitable training set.


This paper is organized as follows: first we present the derivation for a kernel-based regression model capable of predicting response properties simply by letting the response operator act on the kernels.
Next, we implement a representation that allows us to simultaneously train on properties that depend on both the external electric field as well as the internal degrees of freedom of the molecule.
The hydrogen fluoride molecule is used as a toy model to demonstrate the principle.
We benchmark the operator-based machine learning model on a number of existing data sets, that benchmark forces, energies and dipole moments across chemical space, and show how our response model gives an improvement to learning the dipole moment of molecules when compared to conventional kernel-ridge regression models.
Lastly, we discuss how the model naturally couples force and energy predictions with dipole moment predictions, and we show how the response model can directly predict properties related to second order derivatives, including mixed derivatives, such as infrared intensities, harmonic vibrational frequencies, and normal modes.

\section{Theory}


\subsection{Machine learning model}\label{section:ml}
Within kernel-based regression~\cite{MueMikRaeTsuSch01,scholkopf2002learning,Vovk2013,HasTibFri01}, the total potential energy $\mathbf{U}$ of a query molecule $C$ in its electronic ground-state, can be decomposed into a sum of local energies of its $I$ atoms contributions, which are calculated using a basis of kernels:

\begin{align}
	U^{*}_{C} = \sum_{I \in C} U^{*}_\text{local}\left(q^{*}_I\right) 
	=\sum_{I \in i} \sum_{J} \mathpzc{k}\left( q_J, q^{*}_I\right) \alpha_J\label{eq:local_decomposition}
\end{align}
where $J$ is an atomic environment in the basis, $\alpha_J$ is its regression weight, and $q_I$ is the representation of the $I$'th atom in the molecule.

Writing Eq.~\ref{eq:local_decomposition} in matrix form, we have: 
\begin{align}
\mathbf{U} = \mathbf{K}\bm{\alpha}
\end{align}
Note that in contrast to conventional KRR and Gaussian Process Regression (GPR) based QML models~\cite{RaghusReview2016}, this kernel matrix is no longer symmetric since it relies on atomic kernel functions as a basis set. 

In this work, we approximate a response property $\omega$, i.e.~an observable which can be computed by applying a differential operator $\mathcal{O}$ acting 
on the energy $U^{*}$, defined in   Eq.~\ref{eq:local_decomposition}, 
\begin{align}
{\mathbf{\omega}} = \mathcal{O} [\mathbf{U}] \approx  \mathcal{O}[\mathbf{K}]\bm{\alpha}
\end{align}

The set of regression coefficients, $\bm{\alpha}$, can obtained e.g. by minimizing the Lagrangian
\begin{widetext}
\begin{align}\label{eq:lstsq2}
J(\bm{\alpha}) & = & \sum_{\gamma} \beta_{\gamma} \| \mathcal{O}_{\gamma}[\mathbf{U}^{\rm ref}] - \mathcal{O}_{\gamma}[\mathbf{K} \bm{\alpha} ] \|^2_{L_2(\Omega_{\gamma})} \; \;
\equiv \;\; \sum_{\gamma} \beta_{\gamma} \int_{\Omega_{\gamma}} \Big[\mathcal{O}_{\gamma}[\mathbf{U}^{\rm ref}] - \mathcal{O}_{\gamma}[\mathbf{K} \bm{\alpha} ]\Big]^T\Big[\mathcal{O}_{\gamma}[\mathbf{U}^{\rm ref}] - \mathcal{O}_{\gamma}[\mathbf{K} \bm{\alpha} ] \Big] 
\end{align}
\end{widetext}

with respect to $\bm{\alpha}$ over some training set of known values of $\mathcal{O}[\mathbf{U}^{\rm ref}]$. $\Omega_{\gamma}$ is the domain over which the corresponding operator should be minimized, e.g.~all rotational degrees of freedom if the operator acts on a SO(3) group. 
For simplicity we pick $\Omega$ such that $\int_{\Omega} = 1$ for the remainder of this study. 
$\bm{\alpha}$ can be obtained e.g.~by solving the associated normal equations or using an orthogonal factorization such as a QR or a singular-value decomposition (SVD).
The corresponding normal equation (see supplementary materials for derivation) to this problem is given by
\begin{widetext}
\begin{align}
\label{eq:lsqr_alpha}
 \bm{\alpha} & = &  \Big [ \sum_{\gamma} \beta_{\gamma}   \int_{\Omega_{\gamma}} \mathcal{O}_{\gamma}[\mathbf{K}]^T  \mathcal{O}_{\gamma}[\mathbf{K} ] \Big]^{-1}  \Big[ \sum_{\gamma} \beta_{\gamma}  \int_{\Omega_{\gamma}} \mathcal{O}_{\gamma}[\mathbf{U}^{\rm ref}]^T\mathcal{O}_{\gamma}[\mathbf{K}]  \Big]
\end{align}
\end{widetext}

However, solving the normal equations can be numerically unstable since it effectively squares the condition number, i.e.~$\kappa(\mathbf{K}^T\mathbf{K})=\left(\kappa(\mathbf{K})\right)^2$. 

For the practical implementation and the results discussed in the following, an SVD factorization has been used to solve Eq.~\ref{eq:lstsq2}, as it is has several practical and efficient implementations. In contrast to the QR factorization, the SVD factorization is also numerically stable, even if $\mathbf{K}$ is rank-deficient, e.g.~if $\mathbf{K}$ contains rows or columns that corresponding to atoms or molecules that are identical or only differ by symmetry operations to which the representation is invariant.

In the case of under-determined equations, the SVD factorization is performed ignoring singular values smaller than a threshold, which can be treated as a hyperparameter similarly to regularization within ordinary KRR.

\subsection{Operators}\label{section:opperator}


This section is dedicated to discussing some important response operators in quantum mechanics, defining the domain $\Omega$ over which the Lagrangian is to be minimized and to provide corresponding solutions to the integrals in Eq.~\ref{eq:lsqr_alpha}. 



We define the response operator for some external parameter $\vec{\eta} = \{\eta_x,\eta_y,\eta_z\}$ which can be written as $\mathcal{O}_{\delta\vec{\eta}} \equiv \dfrac{\partial}{\partial\vec{\eta}}$. Applying such an operator would map a the scalar field to a three dimensional vector field. All rotational degrees of freedom can then be integrated out with the following solutions.
The solutions to the two integrals in Eq.~\ref{eq:lsqr_alpha}, respectively, are thus

\begin{align}
    \int_{\Omega_{\delta\vec{\eta}}} \mathcal{O}_{\delta\vec{\eta}}[\mathbf{K}]^T  \mathcal{O}_{\delta\vec{\eta}}[\mathbf{K}] &= \dfrac{1}{3}\sum_{\footnotesize {\nu} \in x,y,z} \Big(\dfrac{\partial}{\partial \eta_k} \mathbf{K} \Big)^T\Big(\dfrac{\partial}{\partial \eta_{\nu}} \mathbf{K} \Big) \\
\int_{\Omega_{\delta\vec{\eta}}} \mathcal{O}_{\delta\vec{\eta}}[\mathbf{U}^{\rm ref}]^T\mathcal{O}_{\delta\vec{\eta}}[\mathbf{K}] &= \dfrac{1}{3}\sum_{\footnotesize {\nu} \in x,y,z} \Big(\dfrac{\partial}{\partial \eta_{\nu}} \mathbf{K} \Big)^T\Big(\dfrac{\partial}{\partial \eta_{\nu}} \mathbf{U}^{\rm ref} \Big).
\end{align}

Similarly this procedure can be used to solve the equations for the second order response operator, with respect to two different perturbations $\vec{\eta}$ and $\vec{\eta'}$:

\begin{align}
    \int_{\Omega_{\delta\vec{\eta}\delta\vec{\eta'}}} \mathcal{O}_{\delta\vec{\eta}\delta\vec{\eta'}}[\mathbf{K}]^T  \mathcal{O}_{\delta\vec{\eta}\delta\vec{\eta'}}[\mathbf{K}] &= \nonumber\\
    \dfrac{1}{9}\sum_{\footnotesize {\nu},{\nu}' \in x,y,z}
    \Big(\dfrac{\partial^2}{\partial \eta_{\nu} \partial \eta'_{{\nu}'}} 
    \mathbf{K} \Big)^T\Big(\dfrac{\partial^2}{\partial \eta_{\nu} \partial \eta'_{{\nu}'}} \mathbf{K} \Big) \\
    \int_{\Omega_{\delta\vec{\eta}\delta\vec{\eta'}}} 
    \mathcal{O}_{\delta\vec{\eta}\delta\vec{\eta'}}[\mathbf{U}^{\rm ref}]^T  
    \mathcal{O}_{\delta\vec{\eta}\delta\vec{\eta'}}[\mathbf{K}] &= \nonumber\\
    \dfrac{1}{9}\sum_{\footnotesize {\nu},{\nu}' \in x,y,z}         
    \Big(\dfrac{\partial^2}{\partial \eta_{\nu} \partial \eta'_{{\nu}'}} \mathbf{U}^{\rm ref} \Big)^T
    \Big(\dfrac{\partial^2}{\partial \eta_{\nu} \partial \eta'_{{\nu}'}} \mathbf{K} \Big)
\end{align}
A step-by-step derivation of these equations is given in the supplementary materials.

Now we can explicitly write the matrix elements for the operators investigated within this study. 
In the following, the indices uppercase $I$, $J$, $K$  correspond to atomic centers, and lower-case $i$, $j$ and $k$ correspond to molecules.

The unperturbed kernel corresponds to the energy or identity operator acting on the kernel.
The elements of the unperturbed kernel $\mathbf{K}$ are given as:
\begin{align}
\left(\mathbf{K} \right)_{iJ} = \sum_{I \in i} \mathpzc{k}\left( q_J, q^{*}_I\right)
\end{align}
The kernel elements that correspond the force, i.e.~minus the nuclear gradient operator acting on the kernel are given by:
\begin{align}
   -\frac{\partial}{\partial x_I^{*}}\left(\mathbf{K} \right)_{IJ} = -\sum_{K \in i} \frac{\partial \mathpzc{k}\left( q_J, q^{*}_K\right)}{\partial x_I^{*}} \qquad \text{where} \qquad I \in i
\end{align}
The kernel elements that correspond to the response to the external electric field $\vec{E}$ are given by:
\begin{align}
    \frac{\partial}{\partial E_\nu^{*}}\left(\mathbf{K} \right)_{i_{\nu}J} = \sum_{K \in i} \frac{\partial \mathpzc{k}\left( q_J, q^{*}_K\right)}{\partial E_\nu^{*}} \qquad \text{where} \qquad \nu \in \{x,y,z\}
\end{align}
Similarly, the nuclear Hessian kernel is given by:
\begin{align}
   \frac{\partial^2}{\partial x_{I'}^{*}\partial x_I^{*}}\left(\mathbf{K} \right)_{I'IJ} = \sum_{K \in i} \frac{\partial \mathpzc{k}\left( q_J, q^{*}_K\right)}{\partial x_{I'}^{*}\partial x_I^{*}} \qquad \text{where} \qquad I',I \in i
\end{align}
Lastly, the kernel that yields the dipole derivatives necessary for the infrared intensities is written as the mixed second order derivative,
\begin{align}
   \frac{\partial^2}{\partial E_\nu^{*}\partial x_I^{*}}\left(\mathbf{K} \right)_{i_{\nu}IJ} = \sum_{K \in i} \frac{\partial \mathpzc{k}\left( q_J, q^{*}_K\right)}{\partial E_\nu^{*}\partial x_I^{*}}\nonumber\\
   \text{where}~~I \in i~~ \text{and}~~\nu \in \{x,y,z\}
\end{align}
We are not aware of any other QML model which can account for these effects simultaneously.

\subsection{Comparison to Gaussian Process Regression}
In conventional GPR, the operators (e.g.~derivatives) of the learned function can be included in the training, and the operators are enforced by adding a kernel for each operator of each learned function in the training set.\cite{RasmussenWilliams} E.g.~including the nuclear gradient in addition to the energy will add one additional kernel function for each gradient component in the training set.

Within our formalism, we do not extend the basis by adding additional kernels functions, but we rather enforce the derivatives of the kernel elements in the regression. 
Note that our formalism assigns only one $\alpha$ coefficient per atom, regardless the dimensionality of the perturbation.  
This choice of basis has similarities to the sparsification introduced by Bart\'{o}k and Cs\'{a}ni,\cite{GAPtutorial} although the mathematical origins are different.

In practice this means that the number of kernel function evaluations needed to train the model is reduced drastically. For the examples in this work, memory requirements and training times are reduced by factors of $\sim$10 and $\sim$100, respectively, compared to conventional GPR with the same amount of training data.
The fact that the problem can become over-determined also means that training errors can become substantial. 
Here, we found that in some cases they can even become as large as the test set error.

\subsection{Representation}
In this work we extend the Faber-Christensen-Huang-Lilienfeld (FCHL) representation\cite{faber2017alchemical} to explicitly include the dependence on the variable which can be perturbed, i.e.~an externally applied electric field. This is crucial in order to learn, for example, dipole moments.
The FCHL representation consists of a set of $M$-body expansions $\mathcal{A}_M(I) = \{A_{1}(I),A_{2}(I),A_{3}(I), \dots , A_{M}(I)\}$.
The terms in the many-body expansion correspond to element type, interatomic distances, and interatomic angles, for the one-, two-, and three-body terms, up to order $M$, respectively.

It has previously been shown that the off-set in the learning curve is improved when the two- and three-body terms are multiplied by scaling factors such that features that contribute more to the learned property are weighted higher in the regression.\cite{amons2017}
For energy learning, it was shown that $1/r^n$ and an Axilrod-Teller-Muto term\cite{AxilrodTeller,Muto1943} are suitable scaling factors for the FCHL two- and three-body terms, respectively.

In this paper, we extend the FCHL representation to include a dependence on the external electric field.
Our modified FCHL* representation (denoted by an asterisk) compares the same features as the original formulation (i.e. element type, and interatomic distances and angles), but an extra term is added to the scaling function to emulate the  physics of the the electric-field dependence of the representation, and adjust the weighting accordingly.
The new two-body scaling function (denoted by an asterisk) is given by
\begin{equation}
\xi^{*IJ}_{2} = \xi^{IJ}_{2} - \epsilon (\vec{\mu}_{IJ} \cdot \vec{E})
\end{equation}
where $\xi^{IJ}_{2}$ is the $1/r^n$ scaling function in the original FCHL representation, $\vec{E}$ is the externally applied electric field, and $\vec{\mu}_{IJ}$ is a fictitious dipole arising from fictitious partial charges assigned to the atomic site of the atoms $I$ and $J$, and $\epsilon$ is scaling parameter that balances the two terms in the scaling function.
This parameter was fitted \textit{ad hoc} to $\epsilon = 0.005$ Hartree$^{-1}$ using toy models.
The center-of-nuclear-charge convention is used to define the origin of the coordinate system.
In practice the fictitious partial charges are taken from the Gastieger charge model\cite{GASTEIGER19803219} as implemented in Open Babel.\cite{OpenBabel}
However, we note that the exact values of the fictitious partial charges are unimportant, and any partial charge model could likely be used.
Note that the model does not learn these fictitious partial charges, nor does it use these as a proxy to learn the dipole moment.
The model learns the scalar field of the energy, and the charges merely serve as dummy variabls which enforces the right physical dependence of the kernel elements on the electric field.

The augmented three-body scaling function for atom $I$ interaction with the atoms $J$ and $K$ is similarly given by:
\begin{equation}
\xi^{*IJK}_{3} = \xi^{IJK}_{3} - \epsilon (\vec{\mu}_{IJK} \cdot \vec{E})
\end{equation}
where $\xi^{IJK}_{3}$ is the Axilrod-Teller-Muto scaling factor used to weight the three-body terms in the FCHL representation, and $\vec{\mu}_{IJK}$ is the fictitious dipole arising from fictitious partial charges assigned to the atomic site of the atoms $I$, $J$ and $K$.

In the absence of an externally applied electric field, the FCHL* kernel elements are identical with the original FCHL kernel elements, but the derivative with respect to a perturbing field is now non-zero.
We also note that this representation is "non-polarizable"; the second derivative of the representation with respect to the field is zero with a linear kernel.
This could be amended e.g.~by using on-site multipoles moments with polarizability tensors, e.g.~from a polarizable force field or chemical-potential equalization charge model, rather than a static charge model.

\section{Results}
\subsubsection{Toy model for forces}
In this section we demonstrate numerically the response of the kernel elements with respect to two very different  kinds of perturbations, namely (1) the nuclear coordinates, and (2) an external electric field.
The hydrogen fluoride molecule (H-F) is used as a toy model, and to show how including vector quantities in the training improves learning.

\begin{figure}[!ht]
\centering
 \includegraphics[width=\linewidth]{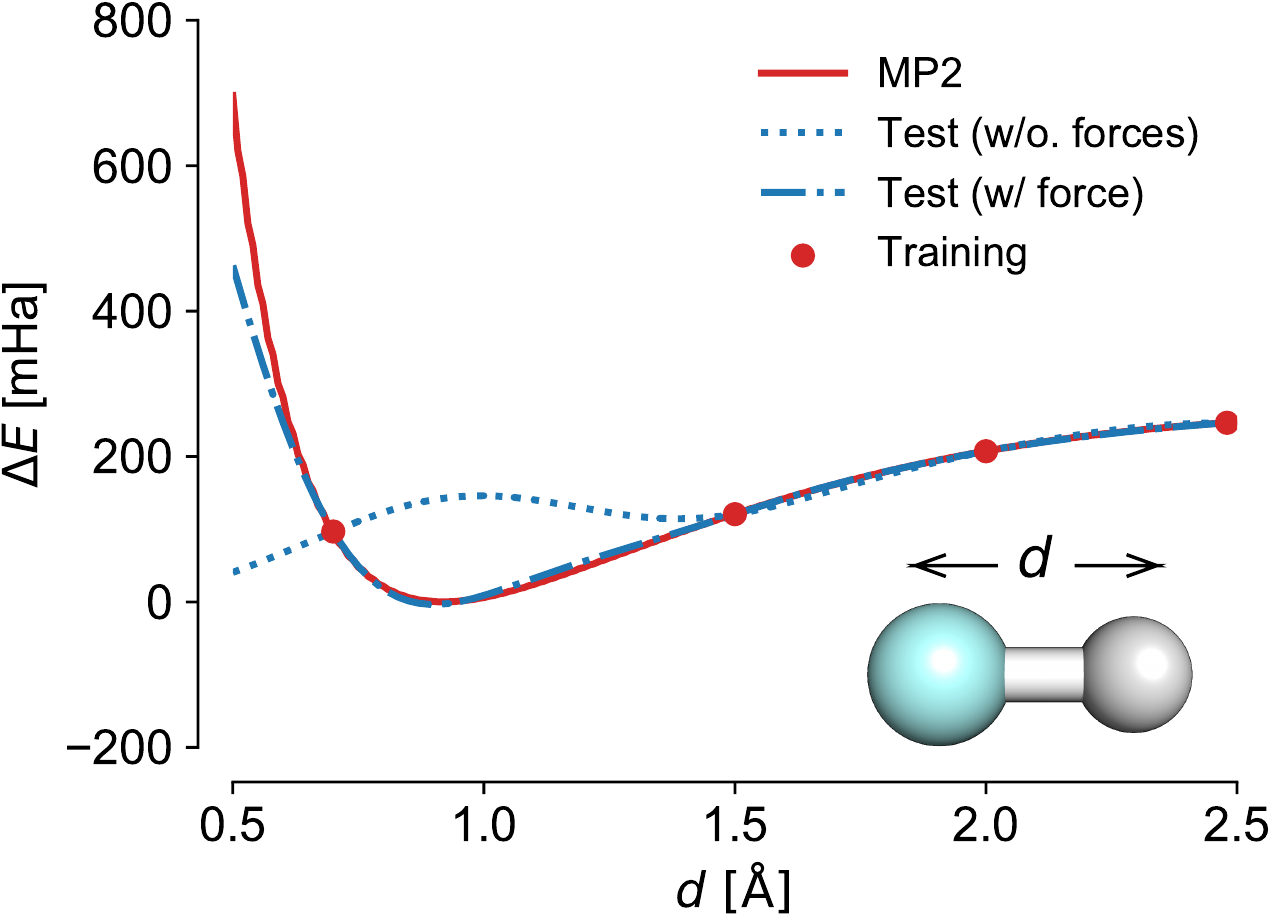}
 \caption{ \label{fig:toy_grad}
The MP2/aug-cc-pVTZ potential energy surface of the hydrogen fluoride (H-F) molecule is displayed as a solid red line.
Four training points (red dots) are selected and two models are trained and used to predict the potential energy surface: One including the interatomic force in addition to the MP2 energy (blue, dash-dotted), and one using only the MP2 energy (blue, dotted).
    }
\end{figure}

We now show how the derivative of the kernel improves learning learning the potential energy of H-F. The MP2/aug-cc-pVTZ potential energy surface for the H-F molecule is used as training data. Selecting four training points (see Fig.~\ref{fig:toy_grad}), models were trained on these four points with and without the interatomic forces in the training set.
Without training on forces using the FCHL representation with the default values, the resulting model describes the dissociation curve poorly; at the minimum-energy distances it even predicts a spurious transition state, and the energy decreases sharply for $r \rightarrow 0$. 
When the forces are included, however, the potential energy surface is reproduced remarkably almost quantitatively, despite only four points being used to fit the model.

\subsubsection{Toy model for electric field}
\begin{figure}[!ht]
\centering
 \includegraphics[width=\linewidth]{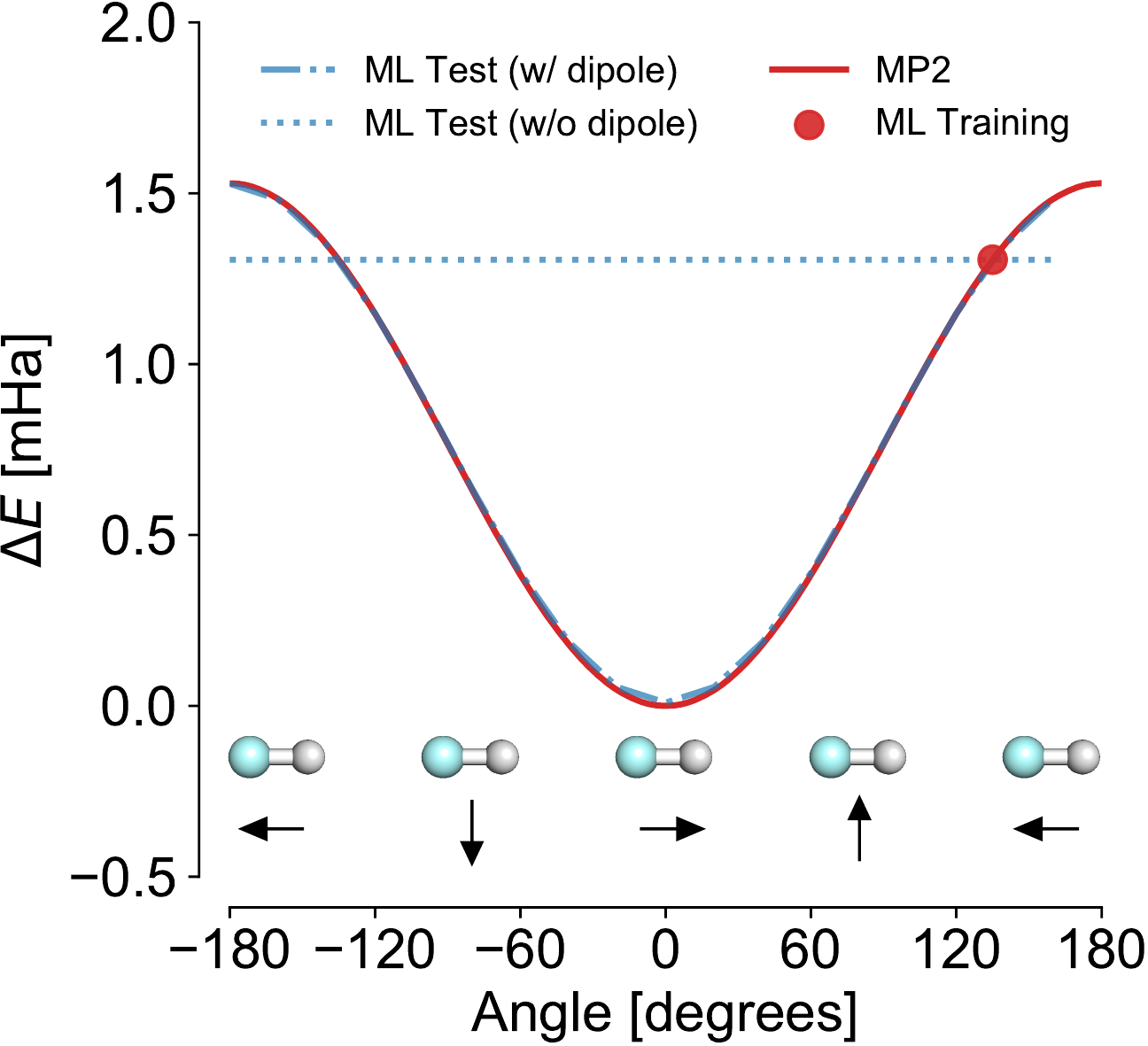}
 \caption{ \label{fig:toy_dipole}
 A hydrogen fluoride (H-F) molecule is placed in an external electric field of 0.001 a.u., and the  MP2/aug-cc-pVTZ energy is calculated as a function of the angle between the H-F molecule and the field vector, displayed as a red line.
 A single point is selected as training point (red dot), and two models are trained, and used to predict the energy in the electric field: one including the dipole moment of the molecule in addition to the MP2 energy (blue, dash-dotted), and one using only the MP2 energy (blue, dotted).
 The alignment between the field and the molecule is sketched at the bottom for clarity.
    }
\end{figure}

Here we demonstrate the effect of including the dipole moment in addition to the energy in the training data.
We now use a GPR model since our approach in section~\ref{section:ml} would only contain two basis functions, while we are including up to four components, i.e.~energy and dipole moment components.
The toy model demonstrates the properties of the FCHL* representations which are fully transferable to the ML approach we present herein.
We place a H-F molecule in an electric field of 0.001 a.u., and which is rotated 360 degrees, and the energy and dipole moment are calculated at each step of 1 degree at the MP2/aug-cc-pVTZ level of theory.
We select just one point in as training set, and train two GPR models, one with the MP2 energy and dipole moment components, and the other with the MP2 energy and without the dipole moment.
The energy prediction of these models as a function of the rotation of the field are displayed in Fig.~\ref{fig:toy_dipole}.
Without fitting to the dipole moment, the energy change due to electric field is close to 0, only fluctuating by a bit of numerical noise from the fit.
When the dipole moment is included, the curve is reproduced almost quantitatively, with only a negligible  deviation at the lowest energy point, presumably due to very small polarization effects and numerical noise.

This demonstrates how including a dipole-like dependence on the electric field in the representation is an efficient way to capture underlying physics of the dipole moment into the kernel.

\subsection{Force and energy learning}
\begin{figure*}[!ht]
\centering
 \includegraphics[width=\linewidth]{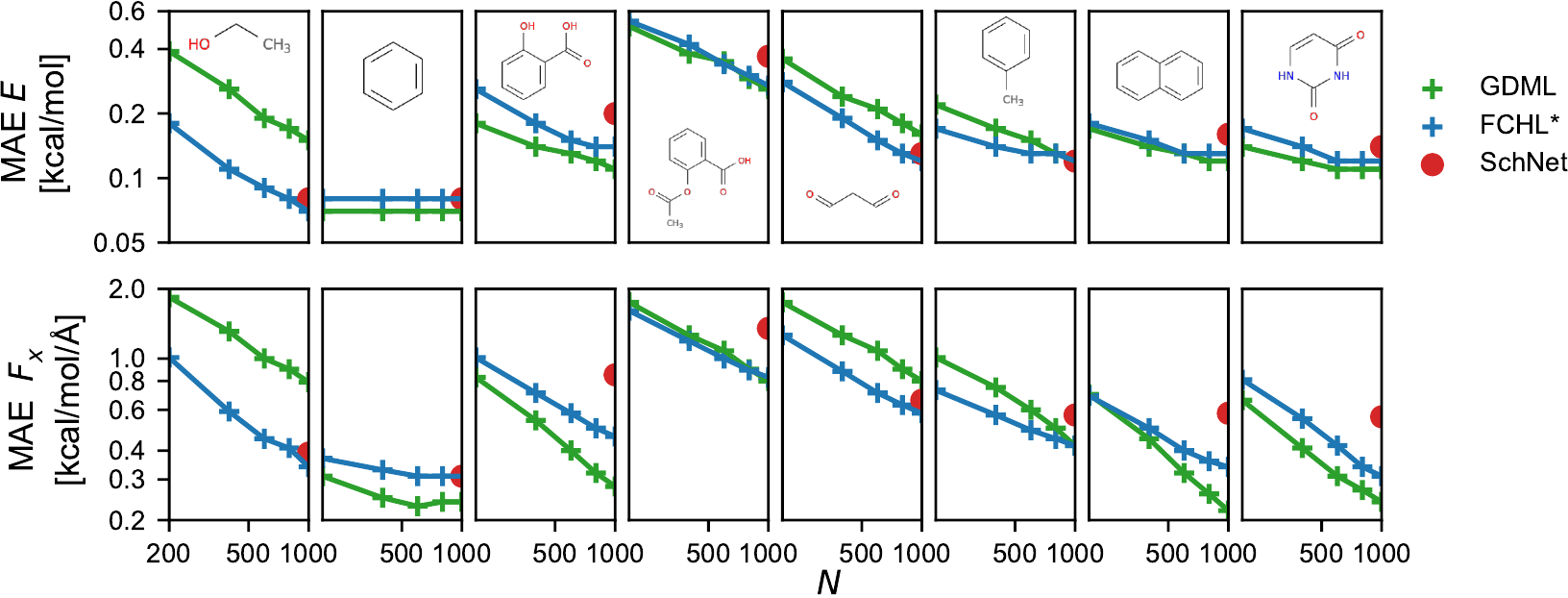}
 \caption{ \label{fig:md17}
 The two figures show the learning curves of our model for the MD17 dataset, for the eight molecules in the MD17 dataset (from left to right) ethanol, benzene, salicylic acid, aspirin, malonaldehyde, toluene, naphthalene, and uracil.
 The out-of-sample MAE energy prediction ($E$, top row) and MAE force component prediction ($F_{X}$,bottom row) is shown for the presented FCHL* (blue) model as well as for the GDML\cite{chmiela2017machine} (green) and SchNet models (red).\cite{schutt2017schnetarxiv,schutt2018schnet}
}
\end{figure*}

\begin{figure}[!ht]
\centering
 \includegraphics[width=\linewidth]{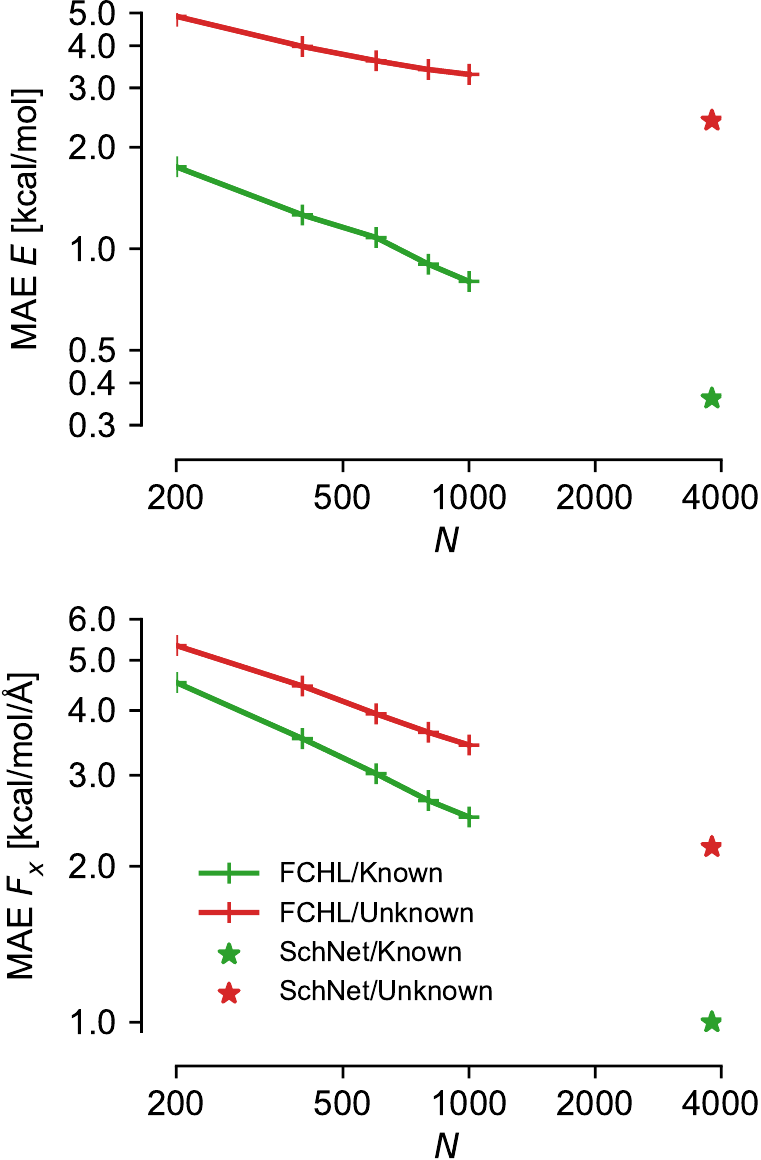}
 \caption{ \label{fig:iso17}
 The two figures show the learning curves of our model for the ISO17 dataset, in addition the accuracy for SchNet when using 4,000 training samples is shown. 
 Left shows the out-of-sample MAE energy prediction for a set of isomers known to the trained machine ("known") and for a set of unknown to the machine ("unknown"). 
 Right shows the out-of-sample  MAE force prediction for the same two sets.
 Note that "known" in this context only concerns whether the isomers are included in the training set or not.
 In both cases only isomers with a conformation unknown to the machine are used to as test data. 
}
\end{figure}

Here we use the FCHL* representation within the presented machine learning algorithm to study two existing benchmark sets for learning forces and energies. 
The MD17 consists of molecular dynamics (MD) snapshots  from MD trajectories of 8 different molecules, for which reference forces and energies are given\cite{chmiela2017machine}.
Similarly, the ISO17 consists of MD snapshots of isomers with the chemical formula C$_7$O$_2$H$_{10}$.
The ISO17 additionally comes with two different test sets\cite{schutt2017schnetarxiv,schutt2018schnet}.
One that consists only of isomers with a connectivity that is present in the training set ("known"), and one that only contains isomers with connectivity that is not present in the training set ("unknown").
Briefly the two datasets benchmark the conformational freedoms and constitutional freedoms of molecules, respectively. Since there is no electric field applied to the molecules in these data sets, note that the FCHL* representation reduces to the original FCHL representation~\cite{FCHL}.

Learning curves for the two datasets are displayed in Figures~\ref{fig:md17} and~\ref{fig:iso17}.
For the MD17 dataset, the out-of-sample MAE errors of predicted energies are similar between FCHL*, GDML and SchNet, with SchNet being slightly less accurate in most cases (See Fig.~\ref{fig:md17}).
FCHL* and SchNet perform best for ethanol and malonhaldehyde, while GDML is the best for salicyclic acid and naphthalene. The case of benzen and uracil is interesting. For benzene, all models show little to no progress  for energies at rather low error values, and the force learning is very weak. Uracil is best modeled by GDML, with relatively poor SchNet forces, and FCHL being in between. 
At this point, we remind the reader that the GDML approach is only applicable to a given system, while 
 FCHL* and SchNet are capable of learning across chemical space. 
 
Performance across constituational space is tested on the constitutional isomers in the ISO17 dataset (Fig.~\ref{fig:iso17}). 
For the two test sets of "known" and "unknown" molecules in the ISO17, the FCHL* model displays a good learning rate, that is qualitatively comparable to the SchNet model.
Note that here, the name "known" only implies that the isomers of the same constitution are known to the machine, but not the conformations in the test set.
Unfortunately the learning curves between the FCHL* models and SchNet do not overlap, so the two models cannot be compared quantitatively here, but the out-of-sample accuracy seems comparable.

Overall, we find that our operator approach leads to forces with state-of-the-art accuracy, on par with two of the most accurate models already published in literature.

\subsection{Learning dipole moments of QM9}
\begin{figure}[!ht]
\centering
  \includegraphics[width=\linewidth]{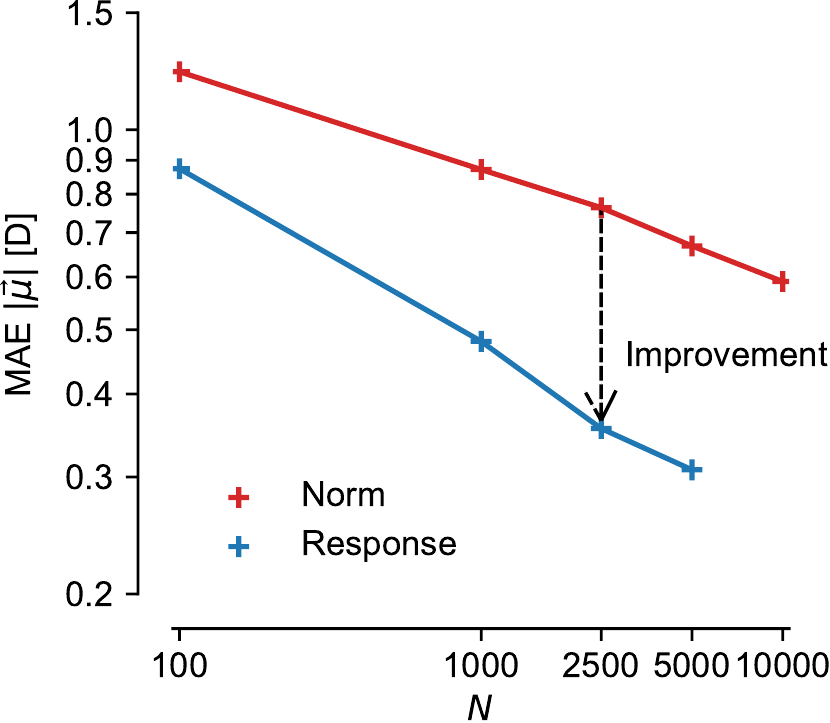}
  \caption{ \label{fig:qm9_dipole}
  The figure displays the out-of-sample prediction error of the dipole norm as a function of QM9 training data set size. 
  The red curve corresponds to a conventional KRR model learning the scalar with the original FCHL representation (taken from Faber \textit{et al.}\cite{FCHL}).
 The blue curve shows the predictions from a machine trained on the energy and dipole moments of QM9 molecules, which in turn predict the dipole vector, from which the norm is calculated.
  }
\end{figure}

Prediction errors of machine learning models of dipole moments converge slowly for conventional QML models~\cite{SingleKernel2015,googlePaper2017,FCHL}.
Here we demonstrate how including the underlying physics for the dipole moment into the representation improves the learning rate, compared to learning the dipole norm with conventional kernel-ridge regression.
We compare two approaches to learn the dipole moment norm of the molecules in QM9; (1) using the  FCHL* representation with the machine learning approach outlined in section~\ref{section:ml} to fit the dipole moments as derivatives of the energy, and (2) simply learning the dipole moment norm as a scalar using kernel-ridge regression with the FCHL representation as done in our earlier paper\cite{FCHL}.
The learning curves of the two models are displayed in Fig.~\ref{fig:qm9_dipole}.
The MAE out-of-sample predicted dipole moment norm is decreased substantially with our new approach.
For instance, training on 5000 random molecules, the out-of-sample MAE error is reduced by 54\% (From 0.67 Debye to 0.31 Debye).
We also note that not only is the learning curve offset lower when the dipole moment operator is used, compared to conventional KRR, but it is also substantially steeper.
This demostrates the strength of the approach of using the correct response operators in the kernel to learn the corresponding response properties.

\subsection{Learning normal modes }\label{section:normal_modes}
In this section we assess the ability of the methodology to predict vibrational normal modes of a number of organic molecules.

We randomly selected 83 molecules from the QM9 dataset with 9 heavy atoms.
For each of these molecules we create a minimal training set, consisting of all sub-fragments of the molecules with up top 7 heavy atoms, following the methodology of Huang and Lilienfeld\cite{amons2017}.
Effectively this approach can be used to prove that the machine can extrapolate from known properties of smaller molecules to predicting the same properties for larger molecules.

For each of the these generated fragments, a conformational search is performed using RDKit\cite{rdkit}, and the unique conformers are minimized at the $\omega$B97xD/6-31G(d) level of theory.
From each of these minimized geometries, a number of distorted geometries are generated using normal mode sampling\cite{smith2017ani} at the same level of theory.
For each of the distorted geometries, a single-point energy and force evaluation is performed at the $\omega$B97xD/6-31G(d) level of theory, and the forces and energies are saved.
Using the sets of distorted fragment geometries for each of the 83 molecules, we train machines with increasing numbers of samples of each fragment in the sets.

\begin{figure}[!ht]
\centering
 \includegraphics[width=\linewidth]{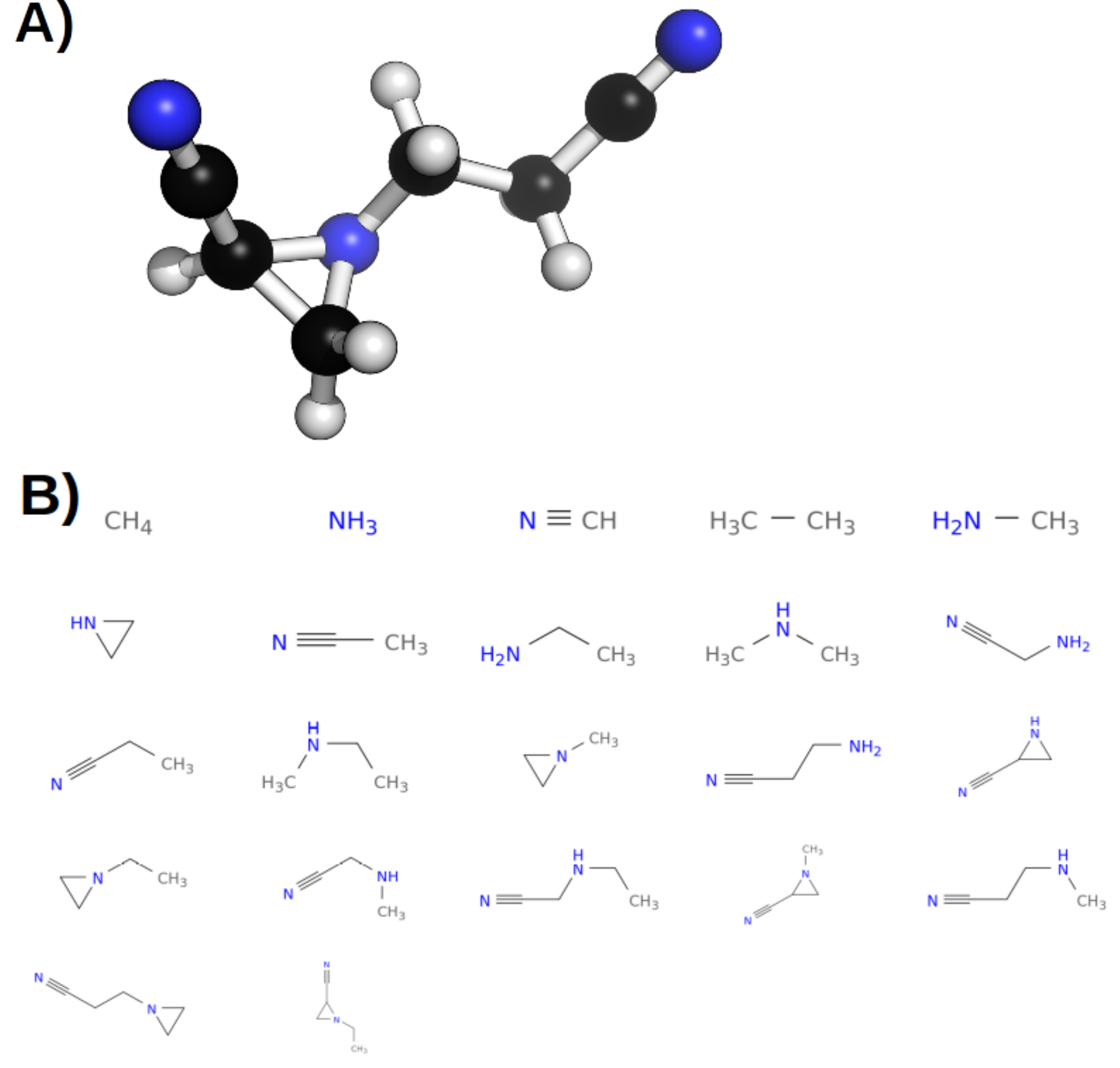}
 \caption{ \label{fig:036682_fragments}
A) displays the QM9 molecule with the ID\# 036682 (SMILES string: \texttt{C1C2C3C4OCOC13C24}) for which normal modes have been predicted in Fig.~\ref{fig:normal_modes_curve}. 
B) displays the fragments identified using the method of Huang and Lilienfeld\cite{amons2017}, which are used to generate the training set for the molecule.
}
\end{figure}

\begin{figure*}[!ht]
\centering
 \includegraphics[width=\linewidth]{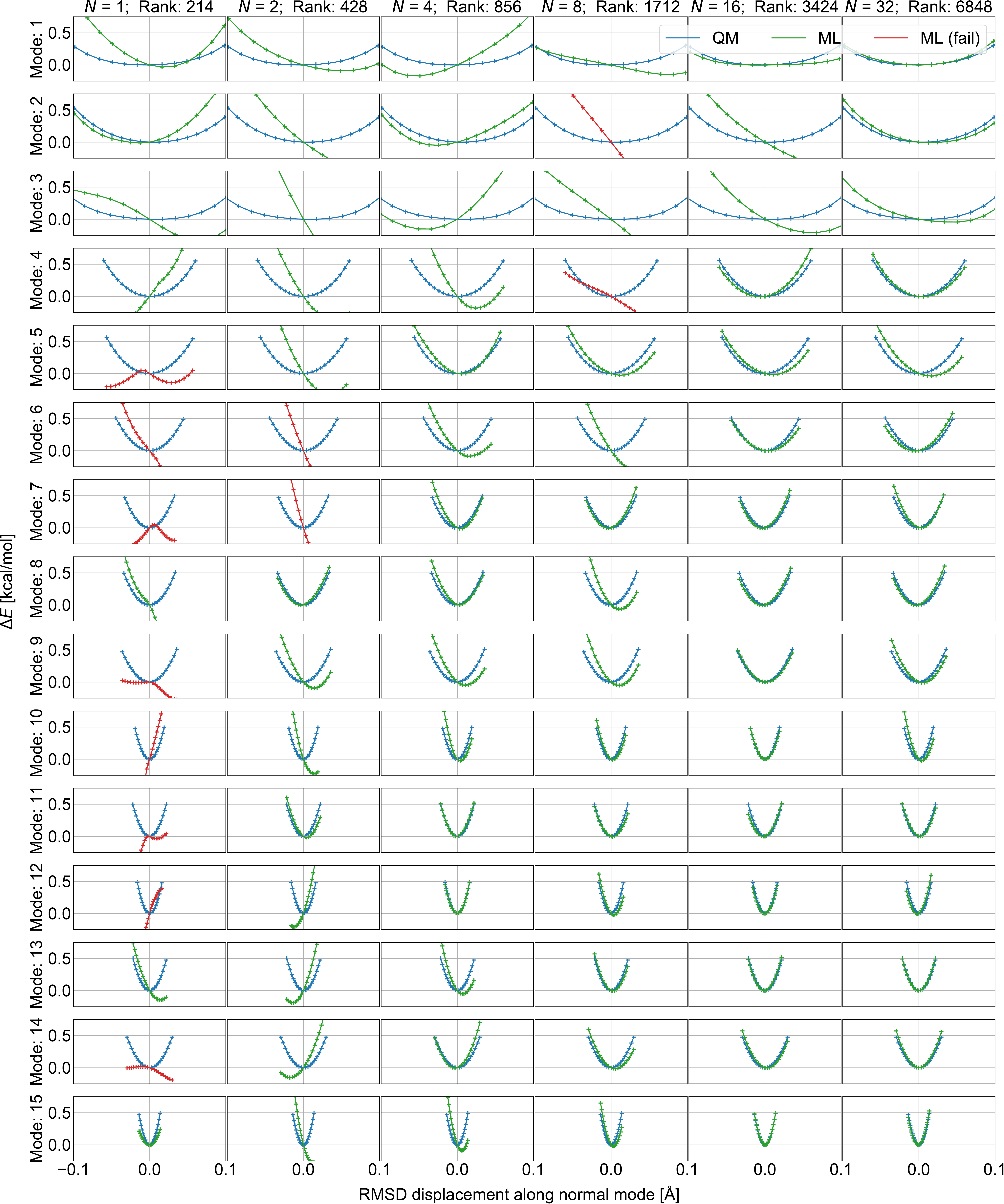}
 \caption{ \label{fig:normal_modes}
ML predicted energy changes of C$_6$N$_3$H$_7$ as a function of distortion 
along each of the 15 normal modes with lowest frequency. 
The molecular structure and its corresponding atom-in-molecule fragments (am-ons) used for training can be seen in Fig.~\ref{fig:036682_fragments}.
Stiffer normal modes are easier to learn and therefore not shown. The complete results set is provided in the SI. 
Each row and column correspond to a normal mode and training set size N/maximum possible rank of kernel matrix, respectively.  
N is the number of samples for each amon (i.e. sub fragment).
Displacement are scaled such that the maximum distortion energy is close to 0.5 kcal/mol.
The X-axis displays the RMSD difference in coordinates to the QM equilibrium geometry, after the molecule has been displaced along that normal mode. 
The Y-axis is the energy difference to the equilibrium geometry, either calculated with QM (blue) or ML (green/red).
The curves predicted from ML are displayed in green if there is a defined minimum within the scan range, and red (fail) otherwise.
}
\end{figure*}

In order to benchmark the performance of the trained machines we set up the following test: 
A vibrational analysis is performed at the  $\omega$B97xD/6-31G(d) level of theory for each of the 83 molecules.
Using the normal modes of the molecules obtained from the vibrational analysis, we generate scans of the potential energy surface along each normal mode.
The scan is consists of structures that are distorted from the equilibrium geometry along each of the normal modes in 10 steps along the positive and negative direction.
The distortions along each normal mode are scaled using the force constants, such that the energy of the geometry with the largest distortion along a normal mode is about 0.5 kcal/mol higher than the equilibrium geometry.
For each of these potential energy scans along the normal modes, we let the trained machines predict the potential energy, and then we compare this to the QM energy.
If the machine predicts a well-defined minimum within the 0.5 kcal/mol scan range, this is counted as a success, otherwise this is counted as a failure.
As an example we show predicted normal modes scans for the 15 normal modes with lowest frequency for a QM9 molecule (C$_6$N$_3$H$_7$, ID\# 036682, SMILES string: \texttt{C1C2C3C4OCOC13C24}) 
in Fig.~\ref{fig:normal_modes}. 
The molecular structure and its corresponding atom-in-molecule fragments (am-ons) used for training can be seen in Fig.~\ref{fig:036682_fragments}.

In addition we present the predictions from machines trained on $N\in\{1,2,4,8,16,32\}$ distorted samples of each sub-fragment in the database.
Data to reconstruct similar plots for all 83 molecules is available from Figshare at \url{dx.doi.org/10.6084/m9.figshare.6994445}.
For the machine trained on only $N=1$ sample per fragment, a total of 11 normal modes do not have a well-defined minimum within the scan range.
By increasing the training set to $N=2$, the machine only predicts two normal modes with minimums outside the scan range. 
At $N=4$, all normal modes have a well-defined minimum inside the scan range, but increasing to $N=8$, two of the low normal modes that corresponds to very non-local conformational changes are not identified correctly to lie within the scan range.
Increasing again to $N=16$ samples, the minimums are well-defined again, and at $N=32$, the QM potential energy curves are almost quantitatively reproduced.

We note that the higher normal modes, which mostly correspond to very local distortions such as e.g. a single hydrogen bond stretching, are almost always very well reproduced.
In contrast, the lower normal modes, which often are more non-local in nature and correspond to very flat energy surfaces require larger training set sizes to reproduce correctly.

\begin{figure}[!ht]
\centering
 \includegraphics[width=\linewidth]{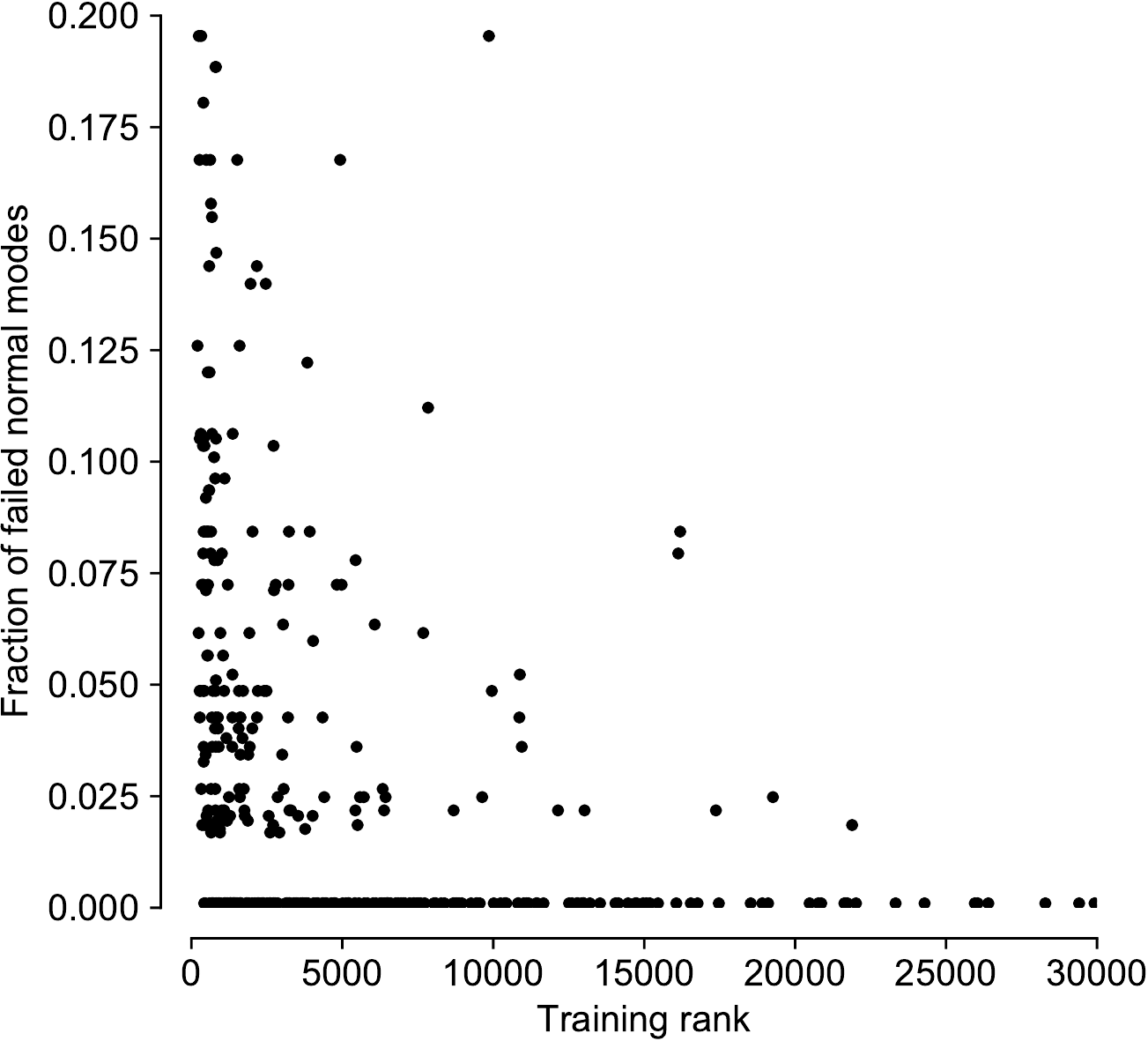}
 \caption{ \label{fig:normal_modes_curve}
Fraction of failed normal mode predictions for 83 QM9 molecules with 9 heavy atoms as a function of training set size.
For each molecule six machines are trained with increasing numbers of molecules in the training set.
The X-axis shows the rank of the kernel matrix (i.e. the number of regression coefficients) for each training set used to train a model for a molecule. The Y-axis shows the fraction of modes for the same molecule which the machine predicts a well-defined minimum within a reasonable distance (see text) from the reference equilibrium geometry.
}
\end{figure}

Repeating the same test for all of the 83 QM9 molecules, we can plot the fraction of normal modes which are incorrectly, as function of the training set size, measured as the maximum possible rank of the kernel matrix, corresponding to the number of regression coefficients.
This is plotted for all 83 molecules in Fig.~\ref{fig:normal_modes_curve} for the corresponding machines training on $N\in\{1,2,4,8,16,32\}$ distorted samples of each sub-fragment.
We note a trend that larger training sizes yield a smaller chance, that the machine fails to identify a well-defined minimum close to the minimum in the reference geometry.

\clearpage

\subsection{Infrared spectrum for Dichloromethane}
In order to demonstrate the utility of the above developments, we have combined them in order to learn and predict IR spectra. More specifically, a vibrational analysis is performed to get the harmonic frequencies and the IR intensities for the dichloromethane molecule.
We train models on distorted geometries of the dicholoromethane molecule for which MP2/def2-TZVP energies, forces, and dipole moments had been calculated previously.
The training set consists of 100 distorted geometries which are generated by normal-mode sampling following the protocol of Smith \textit{et al}.\cite{smith2017ani}
Using the trained model, a vibrational analysis is performed in a standard quantum chemistry package (Gaussian09),\cite{Gaussian09} via an interface to the QML code,\cite{qmlcode2017} which supplies the necessary energies and derivatives to the quantum chemistry program.
As a reference we compare the IR spectrum from the vibrational analysis on potential energy surface of the machine learning model to the IR spectrum from a standard vibrational analysis at the MP2/def2-TZVP level.

\begin{figure}[!ht]
\centering
 \includegraphics[width=\linewidth]{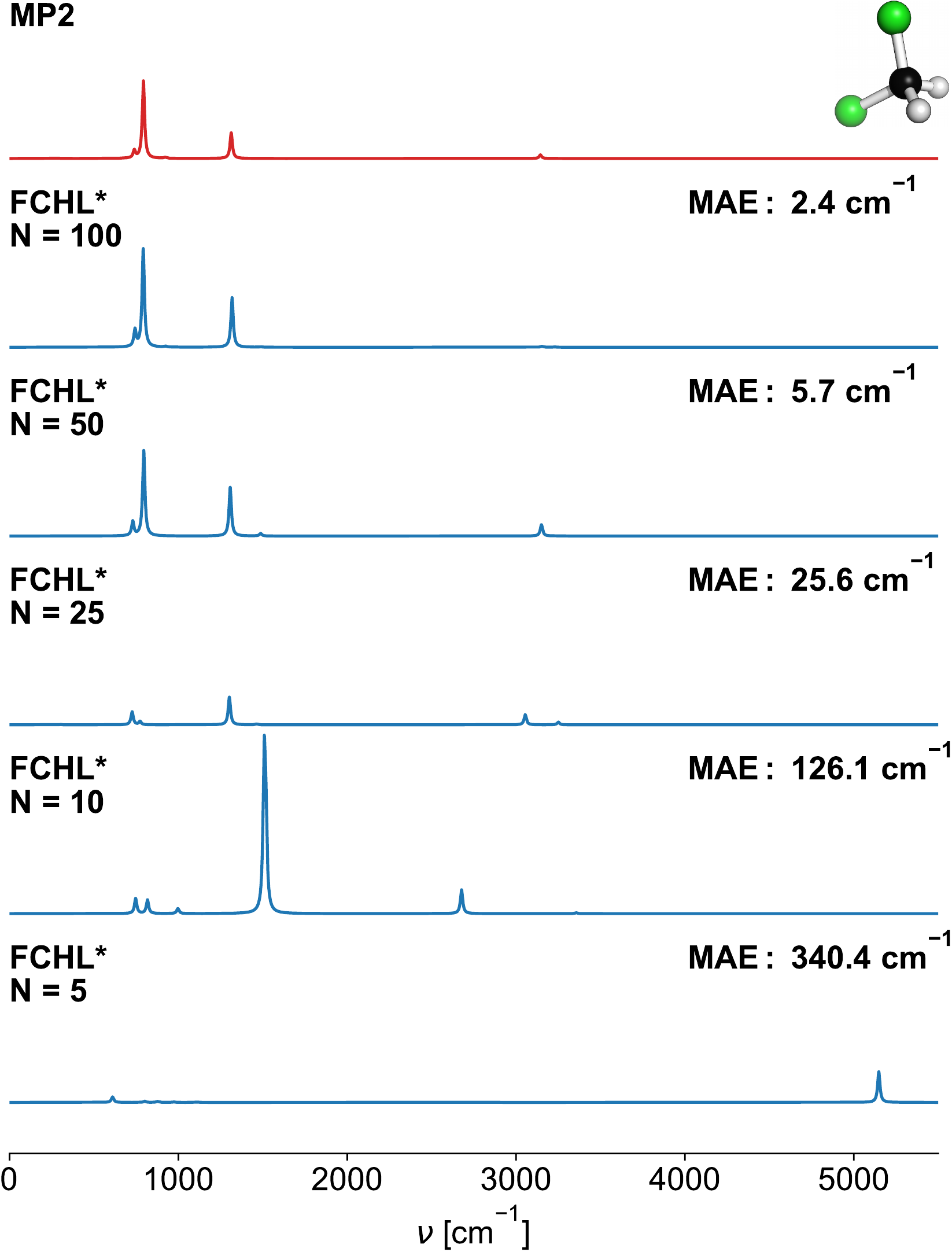}
 \caption{ \label{fig:dichloromethane_ir}
 The unscaled infrared spectrum of dichloromethane calculated via vibrational analysis. 
 (Top/red) calculated at the MP2/def2-TZVP level of theory;
 (bottom/blue) using QML to calculate the necessary derivatives of the energy with respect the nuclear coordinate and the dipole moment. The spectra are convoluted using Lorentzian distributions\cite{Madanakrishna2017} with a width of $\gamma = 8$ cm$^{-1}$.
}
\end{figure}

We have trained 5 models on a decreasing number (100, 50, 25, 10, 5) of randomly selected configurations from the full 100 configurations training set, optimize the geometry and perform the vibrational analysis with each of the trained models.
The resulting IR spectra for dichloromethane are displayed in Fig.~\ref{fig:dichloromethane_ir}.
Qualitatively the FCHL* models reproduce the frequencies of the true MP2 reference with close agreement between the vibrational frequencies of the tallest peaks, even with as few as 10 training samples.
In the spectrum generated using the largest training set (100 samples), the three most intense peaks in the spectrum are located at 743, 793 and 1318 cm$^{-1}$, compared to 740, 793 and 1315 cm$^{-1}$ for the reference MP2 spectrum. 
Training the model on only five randomly selected samples does not lead to a meaningful IR spectrum, however, already with ten instances, decent frequencies and underestimated intensities are obtained for the first two peaks.
We note that the dichloromethane molecules has 9 normal modes, and it is therefore expected that at the very least 9 samples would be necessary to have the minimally required sampling along all the possible normal modes.
Further increasing the training set size to 25 and 50 samples improves the locations of the peaks to MAE vibrational frequencies of 25.6 and 5.7 cm$^{-1}$, respectively.
At 100 training samples the spectrum is almost at spectroscopic precision with an MAE of only 2.5 $^{-1}$.

This demonstrates the generality of the response operator-based machine learning model. The IR intensities correspond to second order mixed derivatives, indicating that the model accounts even for higher order effects after including only energy and first order derivatives. These results suggest that systematic addition of higher order effects will systematically improve performance even further.

\section{Methodology}
\subsection{Used software}
All energy, gradient and dipole-moment calculations for the H-F molecule were performed in ORCA 4.0.1\cite{orca4} at the MP2/aug-cc-pVTZ level of theory with no RI approximation and the \texttt{NoFrozenCore} keyword. 
The relaxed MP2 density was used to calculate the dipole moment as the correct derivative of the energy.

Since only the dipole norms are supplied in with the QM9 dataset\cite{GDB17,DataPaper2014}, the dipole moment vectors of QM9 were re-calculated using ORCA 4.0.1.
To ensure consistency with the B3LYP/6-31G(2df,p) method and basis set used in the original QM9 dataset, the \texttt{B3LYP/G} option was used for the B3LYP functional\cite{B3LYP} and the 6-31G(2df,p) basis set was manually set up to the same contraction coefficients and exponents as used in the original calculations.

Energies, forces and vibrational analyses for the QM9 molecules and fragments in section \ref{section:normal_modes} were calculated at the $\omega$B97xD/6-31G(d) level of theory using the Gaussian09 program.\cite{Gaussian09}
The structures and data corresponding data can be found in comma-separated values format from Figshare at \url{dx.doi.org/10.6084/m9.figshare.7000280}.

The forces, energies and dipole moments of the dicloromethane molecule were calculated at MP2/def2-TZVP level of theory in the Gaussian09 program.
The MP2 vibrational analysis was also carried out in Gaussian09.
The vibrational analyses that uses the machine learning were also carried in Gaussian09 via a Python interface to the machine learning code, and the keywords \texttt{freq=(numer,fourpoint,step=100)} to get the second derivatives.
Our current implementation employs two-point numerical first derivatives, except for geometry optimizations for which it was necessary to use a five-point numerical derivative due to the sensitivity to numerical noise in the optimizer.

The reader can carry out machine learning with the presented algorithms, i.e.~implemented kernel functions, efficient solvers and the FCHL* representation. The necessary code is freely available from our open source machine learning toolkit QML\cite{qmlcode2017} at \url{http://github.com/qmlcode/qml}.

\subsection{Hyperparameters}
All hyper parameters of the FCHL* representation were kept fixed to the same values as those found to be optimal in our previous paper~\cite{FCHL}, and the only new parameter is the newly introduced $\epsilon = 0.0005$ Hartree$^{-1}$ parameter in the scaling functions.
The Gaussian kernel width was set to $\sigma = 0.64$ in all cases, and the cap for smallest singular values to keep in the SVD decomposition was set to $10^{-9}$ in units of the largest singular value.
These parameters were not rigorously fitted to any dataset, so it is possible that more optimal values exist.

\section{Conclusion}
This paper explores a kernel-based supervised machine learning model that is capable of learning response properties by applying the corresponding response operator to the kernel function.
Within this framework, we have extended the FCHL representation by a physically motivated response term for the application of an external electric field.
Using the hydrogen fluoride molecule as toy model, we have demonstrated how the machine learning model and representation can account for the right physics in simple systems with only a minimal number of training samples.
Benchmarking the accuracy of our model for force and energy prediction on the MD17 and ISO17 dataset, our QML models achieve \textit{state-of-the-art} accuracy, on par or better than the GDML and SchNet models.
For learning the dipole norm of the molecules in the QM9 dataset, using the operator formalism leads to an improvement of 54\% compared to learning the same quantity as a scalar with the same representation.
Lastly we allude to the possibility to obtain  higher order derivatives, including mixed derivatives. This idea has been demonstrated by training a model on the energies, forces and dipole moments for the dicholormethane molecule.
Using the resulting model we have performed a vibrational analysis and presented the resulting infrared spectrum which systematically approaches the reference spectrum (calculated at the corresponding \textit{ab initio} level of theory) as more training cases are being added.

Our results suggest that it is advantageous to learn response properties via the corresponding response operators.
The machine learning methodology presented here is, in principle, not limited to derivatives of the energy with respect to the nuclear positions or the external electric field.
We envision to extend the representation to account for a multitude of other properties, such as higher order response properties, as well as magnetic properties such as NMR chemical shifts and spin-spin coupling constants, or alchemical derivatives.
Since the operator formalism is not restricted to any choice of operators it might also be possible to go beyond response operators.
For instance, with the right representation, it should be possible to even learn more fundamental properties of molecules such as the electronic density or the kinetic energy.

\newpage
\begin{widetext}
\section{Supplementary Information for ``Operators in Machine Learning: Response Properties in Chemical Space'': Derivation of operators}

\date{\today}

\maketitle

We predict the total potential energy  $U^{*}_{C}$ of a query molecule $C$ can be decomposed into a sum of local energy contributions which are calculated using a weighted sum of kernels, given in Eq.~\ref{eq:local_decomposition}. The sum runs over the $I$ atomic environments in molecule.

\begin{align}
	U^{*}_{C} = \sum_{I \in C} U^{*}_\text{local}\left(q^{*}_I\right) 
	=\sum_{I \in i} \sum_{J} \mathpzc{k}\left( q_J, q^{*}_I\right) \alpha_J\label{eq:local_decomposition}
\end{align}
A response property, $\omega$, corresponding to the response operator,$\mathcal{O}$ acting on the energy, $U$,  can then be calculated as:
\begin{align}
\label{eq:propperty}
{\mathbf{\omega}} = \mathcal{O} [\mathbf{U}] \approx  \mathcal{O}[\mathbf{K}]\bm{\alpha}
\end{align}
The optimal set of regression coefficients, $\bm{\alpha}$ can be obtained by minimizing the following Lagrangian.
\begin{align} 
J(\bm{\alpha}) =& \sum_{\gamma} \beta_{\gamma} \| \mathcal{O}_{\gamma}(\mathbf{U}^{\rm ref}) - \mathcal{O}_{\gamma}(K \bm{\alpha} ) \|^2_{L_2(\Omega_{\gamma})} \equiv  \\
 \equiv & 
 \sum_{\gamma} \beta_{\gamma} \int_{\Omega_{\gamma}}[\mathcal{O}_{\gamma}(\mathbf{U}^{\rm ref}) - \mathcal{O}_{\gamma}(K \bm{\alpha} )]^T[\mathcal{O}_{\gamma}(\mathbf{U}^{\rm ref}) - \mathcal{O}_{\gamma}(K \bm{\alpha})] = \\
 =& \sum_{\gamma} \beta_{\gamma} \int_{\Omega_{\gamma}}[\mathcal{O}_{\gamma}(\mathbf{U}^{\rm ref})]^T[\mathcal{O}_{\gamma}(\mathbf{U}^{\rm ref})] + [\mathcal{O}_{\gamma}(K \bm{\alpha} )]^T[\mathcal{O}_{\gamma}(K \bm{\alpha} )] - 2[\mathcal{O}_{\gamma}(\mathbf{U}^{\rm ref})]^T[\mathcal{O}_{\gamma}(K \bm{\alpha} )]= \\
 =& \sum_{\gamma} \beta_{\gamma} \int_{\Omega_{\gamma}}[\mathcal{O}_{\gamma}(\mathbf{U}^{\rm ref})]^T[\mathcal{O}_{\gamma}(\mathbf{U}^{\rm ref})] + \bm{\alpha}^T [\mathcal{O}_{\gamma}(K)]^T[\mathcal{O}_{\gamma}(K )]\bm{\alpha}  - 2[\mathcal{O}_{\gamma}(\mathbf{U}^{\rm ref})]^T[\mathcal{O}_{\gamma}(K  )]\bm{\alpha}
\end{align} 
We define the integral over the integration manifold as 1:
\begin{align}
\int_{\Omega_{\gamma}} = 1
\end{align} 
The derivative of the Lagrangian is given by:
\begin{align}
\dfrac{d J (\mathbf{U}^{\rm ref},\bm{\alpha})}{d \bm{\alpha}} = & \sum_{\gamma} \beta_{\gamma} \int_{\Omega_{\gamma}} [\mathcal{O}_{\gamma}(K)]^T  [\mathcal{O}_{\gamma}(K )] \bm{\alpha} +
\bm{\alpha}^T [\mathcal{O}_{\gamma}(K)]^T  [\mathcal{O}_{\gamma}(K )] - 2[\mathcal{O}_{\gamma}(\mathbf{U}^{\rm ref})]^T[\mathcal{O}_{\gamma}(K)] = \\
 =& \sum_{\gamma} \beta_{\gamma} \int_{\Omega_{\gamma}} 2 [\mathcal{O}_{\gamma}(\mathbf{U}^{\rm ref})]^T  [\mathcal{O}_{\gamma}(K )] \bm{\alpha} - 2 [\mathcal{O}_{\gamma}(\mathbf{U}^{\rm ref})]^T[\mathcal{O}_{\gamma}(K)] = \\ 
 =& 2 \sum_{\gamma} \beta_{\gamma}  \big( \int_{\Omega_{\gamma}} [\mathcal{O}_{\gamma}(K)]^T  [\mathcal{O}_{\gamma}(K )] \bm{\alpha} - \int_{\Omega_{\gamma}} 2[\mathcal{O}_{\gamma}(\mathbf{U}^{\rm ref})]^T[\mathcal{O}_{\gamma}(K)] \big )\\ 
 =&  2 \Big( \sum_{\gamma} \beta_{\gamma} \int_{\Omega_{\gamma}} [\mathcal{O}_{\gamma}(K)]^T  [\mathcal{O}_{\gamma}(K )] \bm{\alpha} \Big ) - 2 \Big(\sum_{\gamma} \beta_{\gamma}\int_{\Omega_{\gamma}} 2[\mathcal{O}_{\gamma}(U)]^T[\mathcal{O}_{\gamma}(K)] \Big )
\end{align} 
We now arrive at the corresponding normal-equation solution to the problem:
\begin{align}
0  = & \dfrac{d J(\bm{\alpha})}{d \bm{\alpha}}    \Leftrightarrow \\
   \bm{\alpha} =& \Big( \sum_{\gamma} \beta_{\gamma}   \int_{\Omega_{\gamma}} [\mathcal{O}_{\gamma}(K)]^T  [\mathcal{O}_{\gamma}(K )] \Big)^{-1}  \Big( \sum_{\gamma} \beta_{\gamma}  \int_{\Omega_{\gamma}} [\mathcal{O}_{\gamma}(\mathbf{U}^{\rm ref})]^T[\mathcal{O}_{\gamma}(K)]  \Big)\label{eq:solution}
\end{align}


%

\subsubsection*{First-Order Differential Operators}
Most response operators in chemistry corresponds to the gradient of the energy with respect a change in 3-dimensional variable, $\Vec{\eta}$, such as the nuclear coordinates or an externally applied magnetic or electric field. 
Here we show the solution to any first-order differential operator acting on the energy and kernel.
\\\\The domain of integration, $\Omega$, is the gradient projected on a sphere.
\begin{align}
\label{eq:r}
\Omega = \{\eta_x,\eta_y,\eta_z \in \mathbb{R} | \eta_x^2 +  \eta_x^2 + \eta_x^2 = (4 \pi)^{-1}\}
\end{align}
\\\\First we project the gradient on a spherical coordinate basis, $\Vec{r}$:
\begin{align}
\label{eq:E}
\mathcal{O} = \nabla_{\eta}^{\theta \phi} &= \nabla_{\eta} \cdot \vec{r}
\end{align}
Where the gradient is given by:
\begin{align}
\label{eq:E}
 \nabla_{\eta} = & (\frac{\partial}{\partial \eta_x},\frac{\partial }{\partial \eta_y},\frac{\partial }{\partial \eta_z}) 
\end{align}
and the normal vector of the sphere on which the gradient is projected.
\begin{align}
\label{eq:r}
\Vec{r}(\phi,\theta)  &= (4 \pi)^{-\frac{1}{2}} (\cos(\phi)\sin(\theta),\sin(\phi)\sin(\theta),\cos(\theta))
\end{align}
The integrals in the Lagrangian (Eq.~\ref{eq:solution}) which corresponds to integrating out rotational degrees of freedom are for the left-hand side
\begin{align*}
    & \int_{\Omega} [\mathcal{O}(K)]^T  [\mathcal{O}(K )] =
    \int_{0}^{\pi}\int_{0}^{2 \pi}  [\nabla_{\eta}^{\theta \phi} K ]^T  [\nabla_{\eta}^{\theta \phi} K] \sin{\theta} d\theta d\phi & = \\
     = & \int_{0}^{\pi}\int_{0}^{2 \pi} \Big [\dfrac{\partial K}{\partial \eta_x}\cos(\phi)\sin(\theta) + \dfrac{\partial K}{\partial \eta_y}\sin(\phi)\sin(\theta) +  \dfrac{\partial K}{\partial \eta_z} \cos(\theta) \Big ]^T & \\ 
   &\Big [\dfrac{\partial K}{\partial \eta_x} \cos(\phi)\sin(\theta) + \dfrac{\partial K}{\partial \eta_y}   \sin(\phi)\sin(\theta) +
   \dfrac{\partial K}{\partial \eta_z} \cos(\theta)  \Big] \sin{\theta}  d\theta d\phi &= \\
   = &\dfrac{1}{4 \pi} \int_{0}^{\pi}\int_{0}^{2 \pi} 
   \bigg (
   \dfrac{\partial K}{\partial \eta_x}^T \dfrac{\partial K}{\partial \eta_x} \cos^2(\phi)\sin^2(\theta) +
   \dfrac{\partial K}{\partial \eta_y}^T \dfrac{\partial K}{\partial \eta_y} \sin^2(\phi)\sin^2(\theta) +
   \dfrac{\partial K}{\partial \eta_z}^T \dfrac{\partial K}{\partial \eta_z} \cos^2(\theta) &+\\
   +&(
    \dfrac{\partial K}{\partial \eta_x}^T \dfrac{\partial K}{\partial \eta_y} +
    \dfrac{\partial K}{\partial \eta_y}^T \dfrac{\partial K}{\partial \eta_x}
   ) \cos(\phi)\sin(\phi)\sin^2(\theta)&+\\
   +&(
    \dfrac{\partial K}{\partial \eta_x}^T \dfrac{\partial K}{\partial \eta_z} +
    \dfrac{\partial K}{\partial \eta_z}^T \dfrac{\partial K}{\partial \eta_x}
   ) \cos(\phi)\sin(\theta)\cos(\theta)&+\\
   +&(
    \dfrac{\partial K}{\partial \eta_z}^T\dfrac{\partial K}{\partial \eta_y} +
    \dfrac{\partial K}{\partial \eta_y}^T \dfrac{\partial K}{\partial \eta_z}
   ) \sin(\phi)\sin(\theta)\cos(\theta)\bigg)\sin{\theta} d\theta d\phi &= \\
    = &
    \dfrac{1}{3} 
    \bigg (
    \dfrac{\partial K}{\partial \eta_x}^T \dfrac{\partial K}{\partial \eta_x}  +
    \dfrac{\partial K}{\partial \eta_y}^T \dfrac{\partial K}{\partial \eta_y}  +
    \dfrac{\partial K}{\partial \eta_z}^T \dfrac{\partial K}{\partial \eta_z} 
    \bigg)    &
\end{align*}
and the right-hand side
\begin{align*}
    & \int_{\Omega} [\mathcal{O}(U)]^T[\mathcal{O}(K)] =
    \int_{0}^{\pi}\int_{0}^{2 \pi}  [\nabla_{\eta}^{\theta \phi} U ]^T  [\nabla_{\eta}^{\theta \phi} K] \sin{\theta} d\theta d\phi & = \\
    = &
    \dfrac{1}{3} 
    \bigg (
    \dfrac{\partial U}{\partial \eta_x}^T \dfrac{\partial K}{\partial \eta_x}  +
    \dfrac{\partial U}{\partial \eta_y}^T \dfrac{\partial K}{\partial \eta_y}  +
    \dfrac{\partial U}{\partial \eta_z}^T \dfrac{\partial K}{\partial \eta_z} 
    \bigg)    &
\end{align*}

\subsubsection*{Second-Order Differential Operators}
Similarly, we define a second-order differential operator, $\mathcal{H}$, e.g.~the Hessian of the energy with respect to the nuclear coordinates:
\begin{align}
\mathcal{H} = \nabla_{\eta_1}  \otimes  \nabla_{\eta_2} = 
\begin{bmatrix}
\dfrac{\partial^2}{\partial \eta_{x} \partial \eta'_{x}} & \dfrac{\partial^2}{\partial \eta_{x} \partial \eta'_{y}} & \dfrac{\partial^2}{\partial \eta_{x} \partial \eta'_{z}} \\
\dfrac{\partial^2}{\partial \eta_{y} \partial \eta'_{x}} & \dfrac{\partial^2}{\partial \eta_{y} \partial \eta'_{y}} & \dfrac{\partial^2}{\partial \eta_{y} \partial \eta'_{z}} \\
\dfrac{\partial^2}{\partial \eta_{z} \partial \eta'_{x}} & \dfrac{\partial^2}{\partial \eta_{z} \partial \eta'_{y}} & \dfrac{\partial^2}{\partial \eta_{z} \partial \eta'_{z}}
\end{bmatrix}
\end{align}
We project the operator on the spherical coordinate basis:
\begin{align}
\mathcal{O} = \mathcal{H}^{\theta \phi \theta' \phi' } &= \Vec{r} \cdot \mathcal{H} \cdot \Vec{r}' = \Vec{r}\cdot \nabla_{\eta}  \nabla_{\eta'} \cdot \Vec{r}'
\end{align}
The integrals in the Lagrangian (Eq.~\ref{eq:solution}) which corresponds to integrating out rotational degrees of freedom are for the left-hand side
\begin{align*}
    & \int_{\Omega} [\mathcal{O}(K)]^T  [\mathcal{O}(K )] =
    \dfrac{1}{16 \pi^2}\int_{0}^{\pi}\int_{0}^{2 \pi} \int_{0}^{\pi}\int_{0}^{2 \pi} 
    [H^{\theta \phi \theta' \phi'} K ]^T  
    [H^{\theta \phi \theta' \phi'} K] 
    \sin{\theta} \sin{\theta'} d\theta d\phi d\theta' d\phi' & = \\
   = &\dfrac{1}{12 \pi} \int_{0}^{\pi}\int_{0}^{2 \pi} 
   \bigg ( &\\
   & 
   [\dfrac{\partial}{\partial \eta_x} \nabla_{\eta'} \cdot \Vec{r}' K]^T
   [\dfrac{\partial }{\partial \eta_x} \nabla_{\eta'} \cdot \Vec{r}' K]
   \cos^2(\phi)\sin^2(\theta) & + \\
   + &  
   [\dfrac{\partial }{\partial \eta_y} \nabla_{\eta'} \cdot \Vec{r}' K]^T
   [\dfrac{\partial }{\partial \eta_y} \nabla_{\eta'} \cdot \Vec{r}' K]
   \sin^2(\phi)\sin^2(\theta) & + \\
   + &
   [\dfrac{\partial }{\partial \eta_z} \nabla_{\eta'} \cdot \Vec{r}' K]^T
   [\dfrac{\partial }{\partial \eta_z} \nabla_{\eta'} \cdot \Vec{r}' K]
   \cos^2(\theta) & + \\
   + & (
    [\dfrac{\partial }{\partial \eta_x} \nabla_{\eta'} \cdot \Vec{r}' K]^T [\dfrac{\partial }{\partial \eta_y} \nabla_{\eta'} \cdot \Vec{r}' K] +
    [\dfrac{\partial }{\partial \eta_y} \nabla_{\eta'} \cdot \Vec{r}' K]^T [\dfrac{\partial }{\partial \eta_x} \nabla_{\eta'} \cdot \Vec{r}' K]
   ) \cos(\phi)\sin(\phi)\sin^2(\theta)&+\\
   + & (
    [\dfrac{\partial }{\partial \eta_x} \nabla_{\eta'} \cdot \Vec{r}' K]^T [\dfrac{\partial }{\partial \eta_z} \nabla_{\eta'} \cdot \Vec{r}' K] +
    [\dfrac{\partial }{\partial \eta_z} \nabla_{\eta'} \cdot \Vec{r}' K]^T [\dfrac{\partial }{\partial \eta_x} \nabla_{\eta'} \cdot \Vec{r}' K]
   ) \cos(\phi)\sin(\theta)\cos(\theta)&+\\
   + & (
    [\dfrac{\partial }{\partial \eta_z} \nabla_{\eta'} \cdot \Vec{r}' K]^T
    [\dfrac{\partial }{\partial \eta_y} \nabla_{\eta'} \cdot \Vec{r}' K] +
    [\dfrac{\partial }{\partial \eta_y} \nabla_{\eta'} \cdot \Vec{r}' K]^T
    [\dfrac{\partial }{\partial \eta_z} \nabla_{\eta'} \cdot \Vec{r}' K]
   ) \sin(\phi)\sin(\theta)\cos(\theta)\bigg) & \\ 
   & \sin{\theta} \sin{\theta'} d\theta d\phi d\theta' d\phi'&= \\
    = &
    \int_{0}^{\pi}\int_{0}^{2 \pi}\dfrac{1}{3} 
    \bigg (
    [\dfrac{\partial}{\partial \eta_x} \nabla_{\eta'} \cdot \Vec{r}' K]^T  [\dfrac{\partial }{\partial \eta_x} \nabla_{\eta'} \cdot \Vec{r}' K]  
    +
    [\dfrac{\partial }{\partial \eta_y} \nabla_{\eta'} \cdot \Vec{r}' K]^T
    [\dfrac{\partial }{\partial \eta_y} \nabla_{\eta'} \cdot \Vec{r}' K]  &+\\
    + &
    [\dfrac{\partial }{\partial \eta_z} \nabla_{\eta'} \cdot \Vec{r}' K]^T
    [\dfrac{\partial }{\partial \eta_z} \nabla_{\eta'} \cdot \Vec{r}' K]
    \bigg) \sin{\theta'}d\theta' d\phi'& = \\
   = &\dfrac{1}{9}\sum_{\footnotesize \nu,\nu' \in x,y,z} \Big(\dfrac{\partial^2}{\partial \eta_{\nu} \partial \eta'_{\nu'}} K \Big)^T\Big(\dfrac{\partial^2}{\partial \eta_{\nu} \partial \eta'_{\nu'}} K \Big)   &
\end{align*}
and the right-hand side:
\begin{align*}
    & \int_{\Omega} [\mathcal{O}(U)]^T  [\mathcal{O}(K )] =
    \dfrac{1}{16 \pi^2}\int_{0}^{\pi}\int_{0}^{2 \pi} \int_{0}^{\pi}\int_{0}^{2 \pi} 
    [H^{\theta \phi \theta' \phi'} U ]^T  
    [H^{\theta \phi \theta' \phi'} K] 
    \sin{\theta} \sin{\theta'} d\theta d\phi d\theta' d\phi' & = \\
   = &\dots & = \\
   = &\dfrac{1}{9}\sum_{\footnotesize \nu,\nu' \in x,y,z} \Big(\dfrac{\partial^2}{\partial \eta_{\nu} \partial \eta'_{\nu'}} U \Big)^T\Big(\dfrac{\partial^2}{\partial \eta_{\nu} \partial \eta'_{\nu'}} K \Big)   &
\end{align*}
\end{widetext}

\begin{acknowledgements}
The National Centre of Competence in Research (NCCR) Materials Revolution: Computational Design and Discovery of Novel Materials (MARVEL) of the Swiss National Science Foundation (SNSF) is acknowledged.
This material is based upon work supported by the Air Force Office of Scientific Research, Air Force Material Command, USAF under Grant No. FA9550-15-1-0026.
Some calculations were performed at sciCORE (http://scicore.unibas.ch/) scientific computing core facility at University of Basel.
\end{acknowledgements}
\bibliography{literatur}

\end{document}